\documentclass[conference]{IEEEtran}
\IEEEoverridecommandlockouts
\usepackage{cite}
\usepackage{amsmath,amssymb,amsfonts}
\usepackage{algorithmic}
\usepackage{graphicx}
\usepackage{textcomp}
\usepackage{xcolor}
\usepackage{multirow}
\usepackage{float}

\usepackage[hyphens]{url}

\def\BibTeX{{\rm B\kern-.05em{\sc i\kern-.025em b}\kern-.08em
    T\kern-.1667em\lower.7ex\hbox{E}\kern-.125emX}}
\begin{document}

\title{BLEKeeper: Response Time Behavior Based Man-In-The-Middle Attack Detection}

\author{\IEEEauthorblockN{Muhammed Ali Yurdagul}
\IEEEauthorblockA{\textit{Computer Engineering Dept.} \\
\textit{TOBB-ETU,}
Ankara, Turkey \\
}
\and
\IEEEauthorblockN{Husrev Taha Sencar}
\IEEEauthorblockA{\textit{Qatar Computing Research Institute} \\
\textit{HBKU,} Doha, Qatar \\
}
}

\maketitle

\begin{abstract}
Bluetooth Low Energy (BLE) has become one of the most popular wireless communication protocols and is used in billions of smart devices. 
Despite several security features, the hardware and software limitations of these devices makes them vulnerable to man-in-the-middle (MITM) attacks.
Due to the use of these devices in increasingly diverse and safety-critical applications, the capability to detect MITM attacks has become more critical.
To address this challenge, we propose the use of the response time behavior of a BLE device \textcolor{black}{observed in relation to select read and write operations and introduce an active MITM attack detection system that identifies changes in response time.}
Our measurements on several BLE devices show that their response time behavior exhibits very high regularity, making it a very reliable attack indicator that cannot be concealed by an attacker.
Test results show that our system can very accurately \textcolor{black}{and quickly} detect MITM attacks while requiring a simple learning approach.

\end{abstract}


\section{Introduction}

Bluetooth Low Energy (BLE, also known as Bluetooth Smart) is a wireless protocol designed to support the use of the IoT devices and smart products that we increasingly encounter in every aspect of our lives, from healthcare and fitness to consumer electronics, and from smart homes to industrial automation.
BLE was introduced by Bluetooth Special Interest Group (SIG) in 2010 as part of the Bluetooth 4.0 Core Specification. 
Thanks to its key features like low cost, ease of implementation, fast connection, and low energy requirement, BLE is at the core of several emerging applications \cite{yang2020beyond}. 
It is estimated that 7.5 billion new devices supporting Bluetooth/BLE protocol will be shipped from 2020-2024 \cite{bluetoothsig}.

With the widespread use of the BLE protocol, the number of attacks exploiting security vulnerabilities of BLE devices rapidly increased as well. 
These attacks include device fingerprinting \cite{gu2018bf,martin2019handoff,celosia2019fingerprinting}, accessing and altering sensitive information  \cite{ryan2013bluetooth,wang2020bluedoor,crackle, blurtooth, 255346}, and  disabling devices \cite{garbelini2020sweyntooth}.
In this regard, the man-in-the-middle (MITM) attack poses the most significant threat for BLE security as it allows an attacker to eavesdrop on traffic and modify communicated information. 
As BLE devices are increasingly used as components of safety-critical systems, the implications of an MITM attack go far beyond confidentiality and integrity of data but potentially leads to loss of control over such systems where actual lives may be at stake. 
For example, with an attack on a wearable medical device, such as an oximeter, the measured heart rate of the patient can be intercepted and altered to mimic that of a healthy person, thereby concealing signs of arrhythmia \cite{newaz2020survey}.

\begin{figure}
	\centering
	\includegraphics[width=7cm]{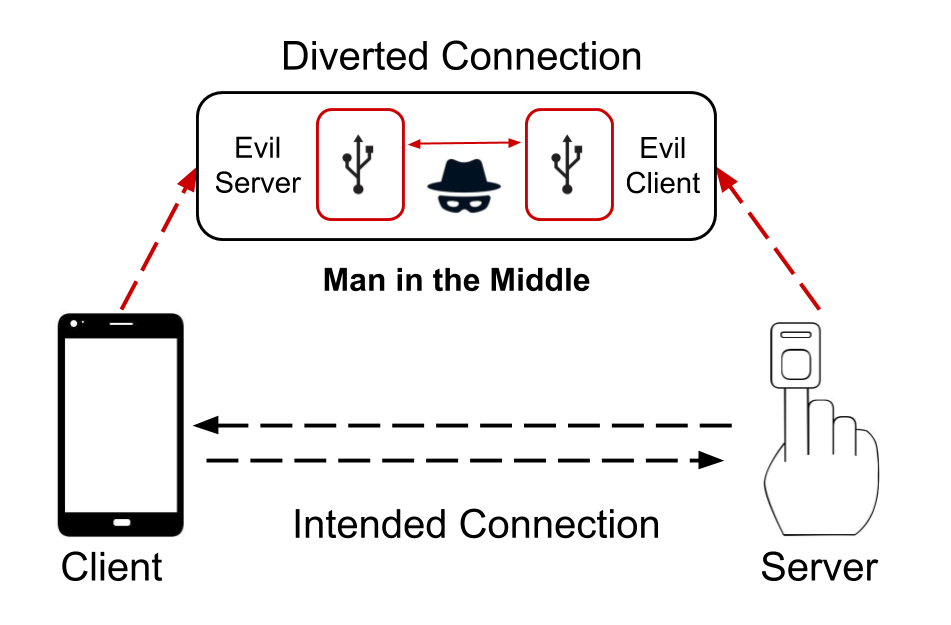}
	\caption{Man-in-the-Middle attack scenario}
	\label{fig:mitmattack}
	\vspace{-0.5cm} 
\end{figure}

As displayed in Fig. \ref{fig:mitmattack}, in a BLE MITM attack scenario, an attacker acts as a client, connects to the server and copies all its properties. The attacker then starts broadcasting spoofed advertising packets, acting as a server and waits for a victim client to connect to it.
Once a victim client, which is unable to differentiate the legitimate server from the spoofed one, establishes a connection with the attacker-operated system, all the communication between the client and server becomes visible to the attacker.
To prevent MITM attacks, Bluetooth standard introduces several pairing mechanisms depending on available hardware capabilities on a device (p. 1641 \cite{bluetoothspec}).
However, devices that do not support any input or output functions, such as wearable medical devices and trackers, use the default pairing method (Just Works) which does not offer any protection against MITM attacks.
It is reported that 71.5\% of BLE devices are under threat of MITM attack because of this limitation and insecure configurations \cite{wen2020firmxray}.
Moreover, it is possible to mount an MITM attack against BLE devices that only support secure pairing methods \cite{blurtooth} \cite{255346}.
In addition, since most BLE devices currently in use are not designed to receive firmware updates, their security posture cannot be changed.
Hence, it can be said that it is impossible to provide protection against MITM attacks on many devices.
\textcolor{black}{This leaves the development of capabilities that can detect the occurrence of MITM attacks as the only viable solution to address this challenge.}

\textcolor{black}{Essentially, detecting an MITM attack requires identifying a characteristic that is intrinsic to a device that when measured exhibits measurable variation from its baseline under an MITM attack and remains invariant to environmental and protocol-related factors.}
With this objective, several MITM attack indicators that relate to  advertising intervals \cite{yaseen2019marc}, received signal strength indicator (RSSI) level,\cite{wublueshield}, carrier frequency offset (CFO) \cite{wublueshield}, power spectral density \cite{galtier2020psd}, and differences in packet interarrival times \cite{lahmadi2020mitm} of a device are proposed.
\textcolor{black}{
Another related characteristic that can be used for detecting MITM attacks is the response time of a BLE device, which is measured as the time it takes for a server to respond to requests made by a client.
In \cite{gu2018bf}, Gu {\em et al.} considered using response time behavior as part of a BLE device fingerprint. 
To further improve on this work, we propose using response time behavior exhibited in relation to specific read and write operations as an attack indicator and introduce an active BLE MITM attack detection system, {\em BLEKeeper}, to more reliably identify attackers.}

Since the interception and forwarding of a data packet sent from the client to the server and vice versa involves additional processing by the attacker in the middle, observed response time from the standpoint of a client device is bound to increase. 
Further, since a device's response time is primarily determined by the hardware and software architecture of a device and depends neither on its mobility nor on an easy-to-spoof protocol level property, it serves as a much more reliable indicator for the occurrence of an MITM attack. 
Moreover, such deviations can be easily detected within a simple hypothesis-testing framework and detection does not require a complex learning approach. 
Our measurements on seven different BLE devices show that these devices exhibit a very consistent response time behavior.
Tests performed using publicly available automated MITM attack frameworks \cite{btlejuice, mirage} show that BLEKeeper can detect MITM attacks with an average accuracy of 98\%.

Next, we provide a brief background about BLE protocol (§ II) and describe the operation of our system (§ III). This is followed by our measurements and evaluation of performance (§ IV). We then conclude our paper with a discussion of our findings (§ V).

\section{Background}

\subsection{BLE Protocol}

The BLE protocol emerged as a variation of the Classic Bluetooth protocol with a consideration for low-power, low-cost devices.  
The protocol runs on the license-free 2.4 GHz ISM band and operates over 40 link-layer channels, each with a 2 MHz bandwidth. 
Three of these channels are designated as advertising channels that are used for device discovery and connection establishment. 
Server devices broadcast their existence to client devices through advertising packets sent over these channels.
The remainder of the channels are data channels and used for communication between devices after a connection is established. 
The process and the format for exchanging device specific attributes over a BLE connection is specified by the Generic Attribute Profile (GATT) which organizes the information in the form of services, characteristics, and characteristic descriptors. 
Services and their characteristics are identified by a universally unique identifier (UUID) value.

\begin{figure}[htp]
\centering
\includegraphics[width=3.6cm]{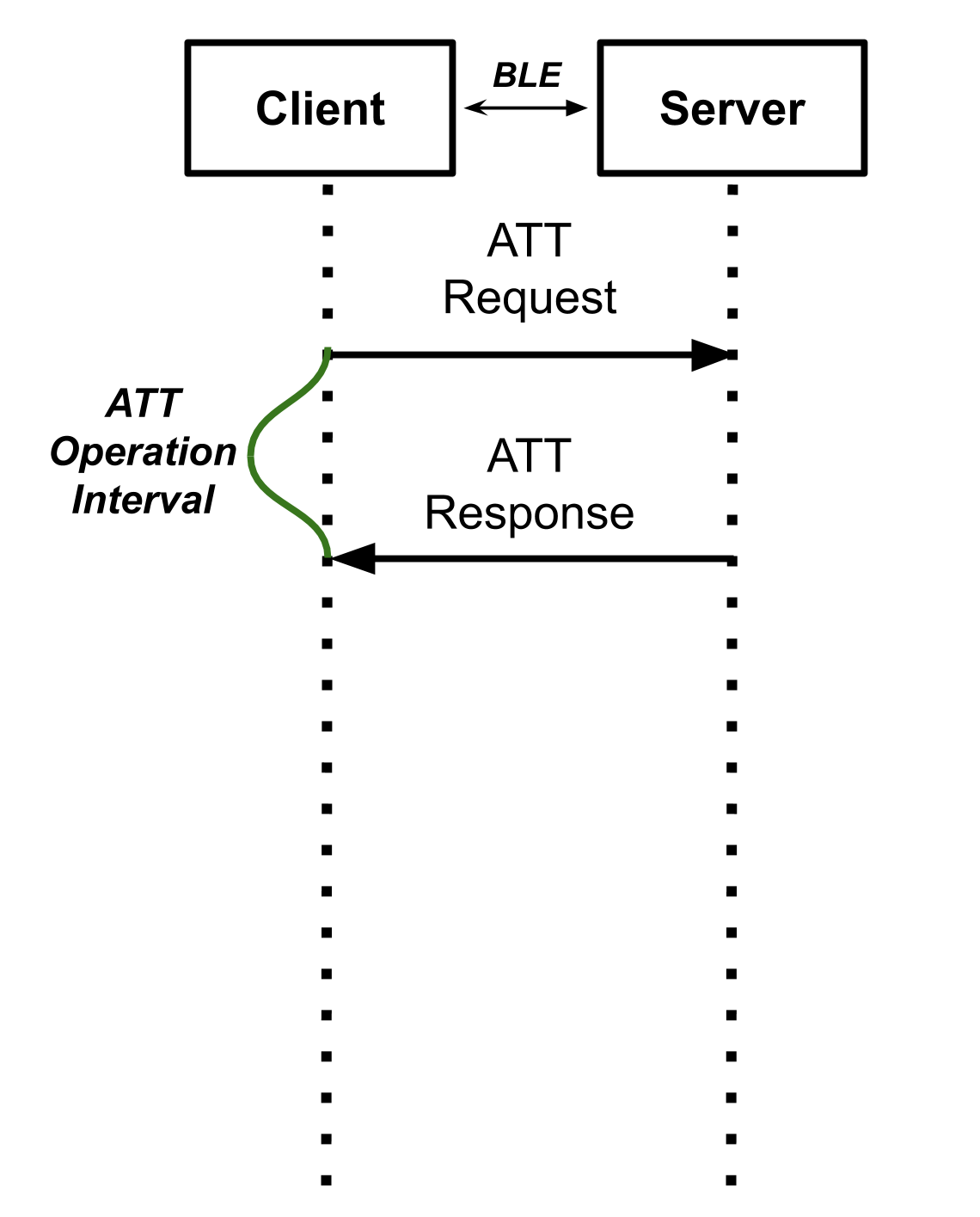}
\caption{Intended ATT request-response traffic.}
\label{fig:legitimateattrequest}
\end{figure}

Characteristics are the main data transfer units communicated between BLE devices, and they consist of properties, values and optional descriptor fields. GATT uses Attribute Protocol (ATT) as the data transfer protocol for discovering services and carrying out operations on characteristics. Permitted operations on a characteristic, such as read, write, notify, and indicate are set as part of its properties (p. 1552 \cite{bluetoothspec}). In this context, indicate and notify operations are both used to notify the client when the value of a characteristic changes with the difference that the indicate operation requires a confirmation from the client. 
Similarly, a read operation obtains a characteristic's value, and a write operation changes it. 
When reading from or writing to a target characteristic value, the client sends a request to the server with the handle of the characteristic that is of interest, as well as the corresponding value if it is a write request. 
The server then responds to each request with a message including the read value or the status of the write request unless otherwise specified (such as, write without response), as demonstrated in Fig. \ref{fig:legitimateattrequest}.

\subsection{Detecting BLE MITM Attacks}

\textcolor{black}{Proposed MITM attack detection approaches rely on some prior knowledge of the BLE devices to identify discrepancies and inconsistencies in device features and behavior when distinguishing connections established with spoofed BLE devices.}
At the simplest, this includes matching the name of a device with its known public MAC address used during advertising \cite{yaseen2019marc}.
However, since attackers can easily spoof both the MAC address and the device name, this association is of limited value.
Another feature that relies on MAC addresses concerns devices that send scan request packets to other devices. 
A scan request packet is aimed at discovering the available service characteristics of a device, and requests from non-whitelisted devices can be identified as leading to an attack \cite{yaseen2019marc}.
But, this control is also ineffective due to address spoofing.

The advertising interval of a BLE device has been another indicator used for identifying spoofed devices\cite{yaseen2019marc}.
Although a change in the rate of advertising packets can be easily determined, the attacker can set its advertising interval to match that of the impersonated device. 
A more reliable indicator proposed for attack detection is the change in the observed RF signal strength, {\em i.e.,} RSSI level, of a BLE device \cite{yaseen2019marc}.
The attacker may spoof the RSSI level in one advertising channel; therefore, monitoring three channels simultaneously can eliminate such an attempt \cite{wublueshield}.
A downside of the use of RSSI is that it mainly applies to stationary devices and does not offer a solution for devices with highly variable RSSI levels such as wearable medical devices and fitness trackers, which comprise a third of all the BLE devices \cite{bluetoothsig}.

Similar to RSSI, power spectral density of BLE signals is also considered for use in detection of MITM attack on BLE devices.
However, this measure has also been tested in static environments, and it requires the device profile to be constantly updated for devices in motion.
One of the most reliable indicators used for detecting MITM attacks is the carrier frequency offset (CFO) which serves a device fingerprint.
The CFO does not depend on the distance between two devices, but has to be estimated at a higher frequency resolution to be accurate, requiring more precise measurements using more advanced sniffers \cite{wublueshield}.

In addition to the above indicators obtained by monitoring the advertising channels, the traffic in data channels can also be used to extract other indicators.
In this regard, Lahmadi {\em et al.} \cite{lahmadi2020mitm} proposed features based on packet interarrival times, which are the timing differences between two subsequent packets, along with RSSI, channel numbers, and distance features to detect anomalies in the communication pattern of a device. A common problem with such learning approaches is the generalizability of performance to different test settings and devices.

A more invariant characteristic of an MITM attack is the change in the response time of a server due to the delay added by the MITM attacker that intercepts, processes, and re-transmits each packet communicated between two parties. 
It has already been shown that anomalies in timing features of messages exchanged between two parties can be used for detecting MITM attacks over TCP \cite{vallivaara2014detecting} and SSL/TLS \cite{kang2018trusted} connections. 
More relevantly, it is shown in \cite{gu2018bf} that response time when coupled with other timing characteristics such as packet interarrival times and advertising intervals, can be used to fingerprint and classify BLE devices.
This work has essentially demonstrated that a device's response time varies depending on the type of ATT operation performed on it, but despite this variation there are discernible differences in response time behavior across devices.

Our approach differs from this study in that we use response time behavior associated with specific operations.
This allows us to obtain a very well-defined behavior that can be used to identify a spoofed BLE device through deviations from its baseline behavior. 
However, since we do not profile the overall response time behavior of a device considering all possible operations, a passive, monitoring-based attack detection approach is not suitable because the attacker may not perform those operations. 
Therefore, BLEKeeper takes a proactive approach and establishes very short-lived connections to BLE server devices in its environment to identify potential MITM attackers that impersonate other BLE devices.
Ultimately, viability of this approach depends on two factors: (i) the degree of variation in BLE device response times and the architecture of the MITM attack tools and how they contribute to device-level variation.
Our detection system BLEKeeper is designed by taking both these factors into consideration.

\section{System Design}

\subsection{Threat Model \& Assumptions}
\label{skilledhacker}

We consider a skilled attacker that uses an automated attack framework based on widely available generic Bluetooth hardware. 
The attacker can not only spoof all protocol-level attributes  of the target device ({\em e.g.,} MAC address and device name) but can also mimic its advertising interval, through its configuration.  
In addition, due to the distance limitation of deployed packet sniffers, we assume our attack detection system is deployed in an enclosed space, like a home, office, or hospital, and outdoor BLE traffic is not considered.
Correspondingly, target BLE devices include those that are commonly used in such environments. 
Finally, we assume both the attacker and the BLE devices can be in motion in the environment where the MITM attack is taking place.

\subsection{System Architecture}

The system consists of two modules, namely, the traffic analyzer module and the attack detection module. 
The first module includes BLE sniffers, needed to capture BLE traffic in the air and a computational unit to run the detection method.
This module continuously monitors all advertising channels and tracks newly established connections in the respective data channels.
BLEKeeper uses commodity hardware in its design to ensure its scalability.
This module essentially collects BLE data packets and obtains response time measurements associated with a BLE device.
This is realized by recording timestamps of all requests sent to BLE devices to read or write GATT characteristic values and the responses generated 
to fulfill these requests.
Then the difference between timestamps of each request-response interaction is computed.

The second module includes three analysis components needed to detect attacks and uses a BLE interface to establish BLE connections. 
The first component creates a profile of each BLE device introduced to the system.  
BLEKeeper can learn the response time behavior of a device in an offline setting by sending several read and write requests to the device or in an online setting by observing interactions between devices as they normally operate.
The second component identifies devices that are suspected of being spoofed.
Then the third component establishes a connection to those devices and performs one or more read and/or write requests to previously specified characteristics.
The traffic analyzer module, which follows the BLE connections to obtain response time measurements, feeds these measurements to the change detection component.
By using the input from the device profiler, this last component makes a decision through a comparison of obtained response time measurement(s) with the known profile of the device in a probabilistic manner to identify anomalous deviations. 
If a spoofing attempt is deemed to be detected, it alerts the operator.

\begin{figure}[!h]
	\centering
	\includegraphics[width=8cm]{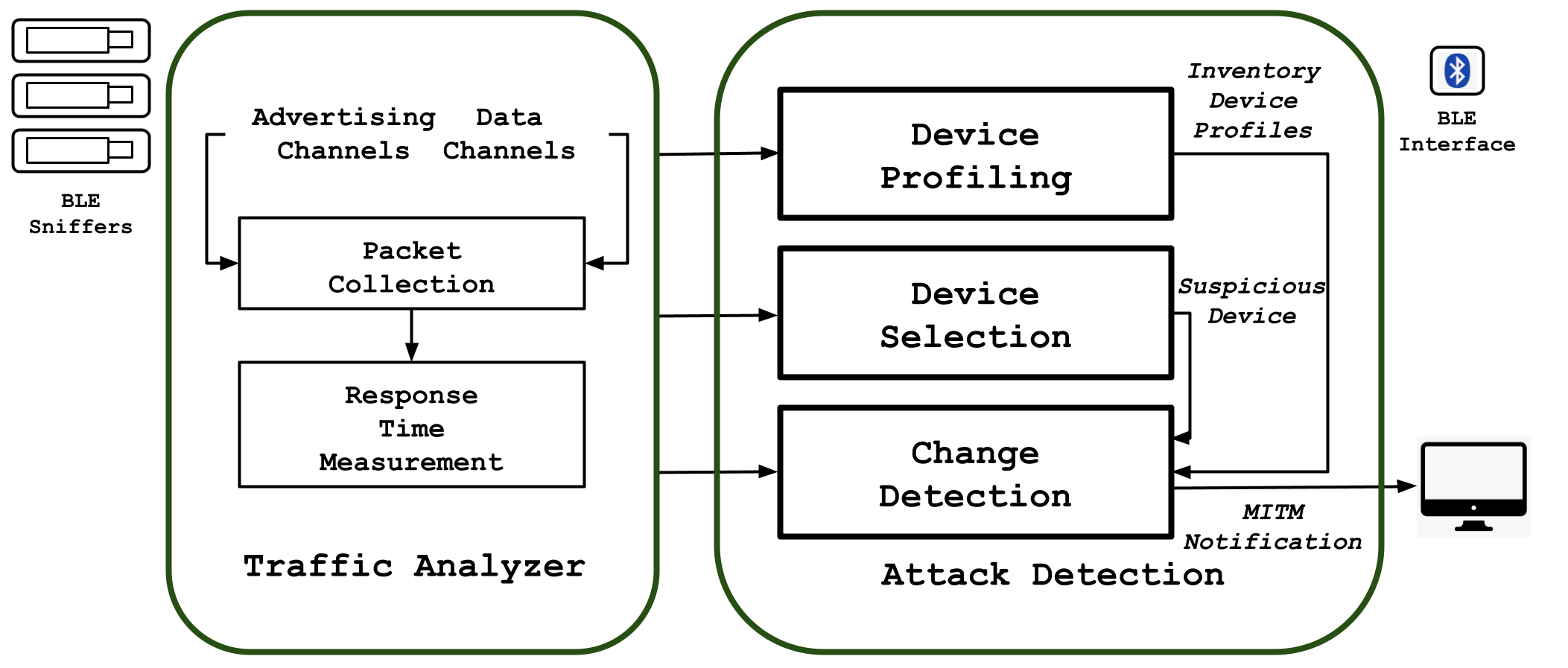}
	\caption{Architecture of the BLEKeeper}
	\label{fig:architecture}
\end{figure}

\subsubsection{Traffic Analyzer}

To accurately measure the response time of a device during profiling or when making a change detection decision, the timing information for the request and response packets must be correctly captured by tracking a connection. 
(We must note here that packet arrival and departure times can also be measured at the BLE interface, however, due to several sources of delay in a system those measurements will not be as accurate as those obtained through sniffing packets over the air.)
Since in the BLE protocol a device can use three channels 
for advertising, a connection request packet can be sent to the server through any of them.
Hence, in order to track a connection which takes place over data channels, this initial packet must be captured. 
Therefore, traffic analyzer deploys three sniffers to be able to capture all packets in the advertising channels. 
When a connection has to be tracked, one of these sniffers is used to capture packet traffic on corresponding data channels. 

\subsubsection{Device Profiling}
BLEKeeper creates a profile of each device to describe its response time behavior. 
The profile $P_{i}$ for a BLE device indexed by the identifier $i$ is defined in terms of the variation in elapsed time between a read or write request sent to a BLE device and the response it generated after processing the request, as shown in Fig. \ref{fig:legitimateattrequest}. 
This variation can be represented using a non-parametric model describing the distribution of response time values, which can be obtained empirically by computing the histogram of measured values.
Considering the response time $\mathbf{Tr}$ to be a random variable, each device profile can be represented by a probability distribution function, $P_i: f^{i}_{\mathbf{Tr}}(Tr)$ that shows the likelihood of observing a particular response time value $Tr$ when using the $i^{t}$ BLE device.

Before the profiling process, a connection to the device is established and the GATT characteristics of the device are determined. As it is generally readable, the characteristic with uuid \texttt{2a00} (Device Name) \cite{assigned} is selected as the readable characteristic. 
In a similar manner, a write request containing \texttt{00} value is sent to a characteristic that has a write property. If a write response is received, this characteristic is used as the writable one.
Then multiple read and write operations are performed on these characteristic values for BLEKeeper to obtain necessary response time measurements. 
Each profile is indexed by an identifier obtained by combining the device name and the MAC address seen in the advertising packets. 
Since the MITM attacker has to spoof both attributes in its advertising packets to avoid detection, the change detection decision will be performed with respect to the device profile created during the initial profiling process.  

\subsubsection{Device Selection}
To identify potential attackers, BLEKeeper keeps a record of all connection requests and preceding advertising packets.
Since an attacker has to first establish a connection with a server to launch an MITM attack, sources of all advertising packets 
that previously established a connection are labeled suspicious as they might have been spoofed.
Only these devices are tested for their response time behavior.

\begin{figure}[htp]
\centering
\begin{minipage}{0.18\textwidth}
\includegraphics[width=\textwidth]{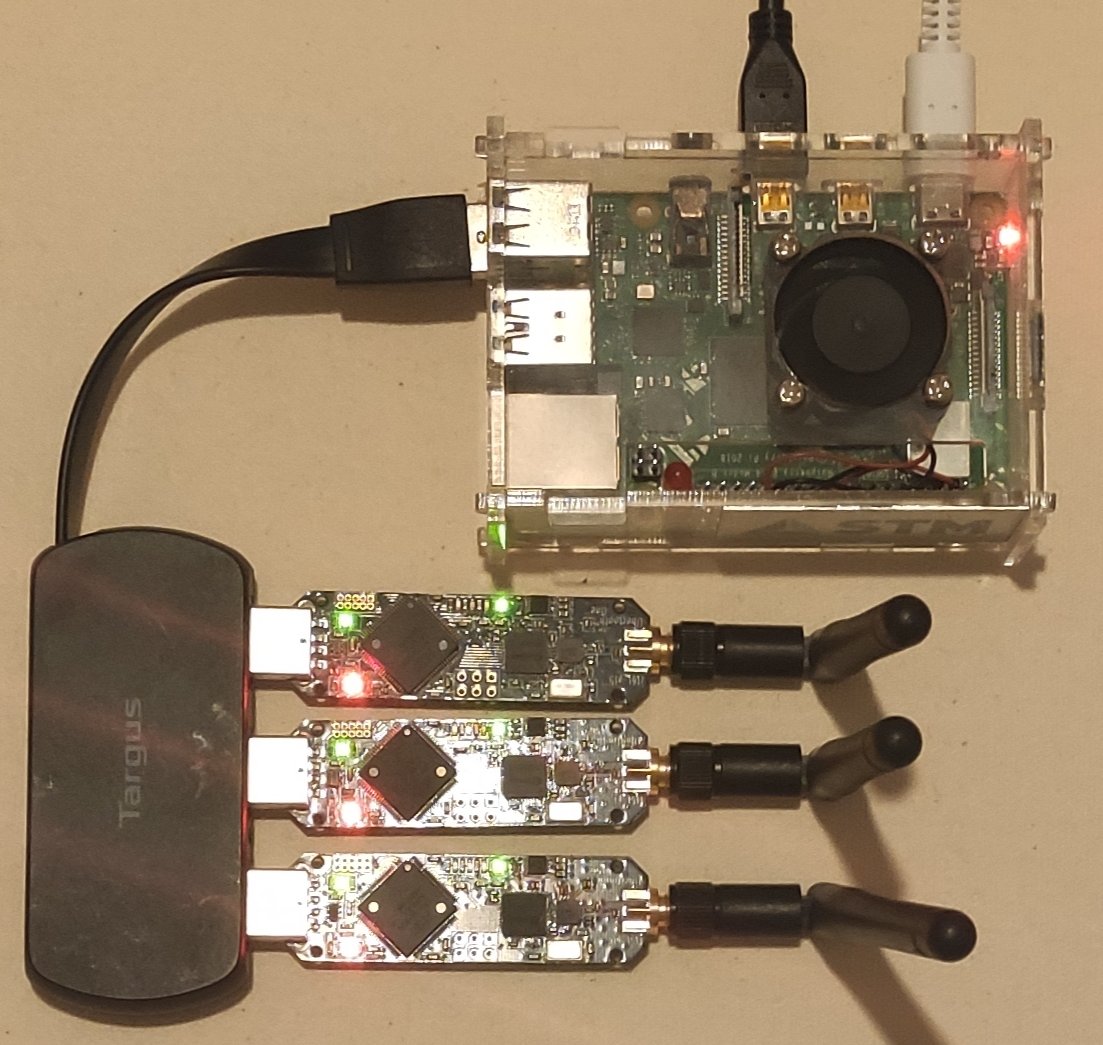}
\caption{A prototype of the BLEKeeper}
\label{fig:detectionsystem}
\end{minipage}
\hspace{0.3cm}
\begin{minipage}{0.23\textwidth}
\includegraphics[width=\textwidth]{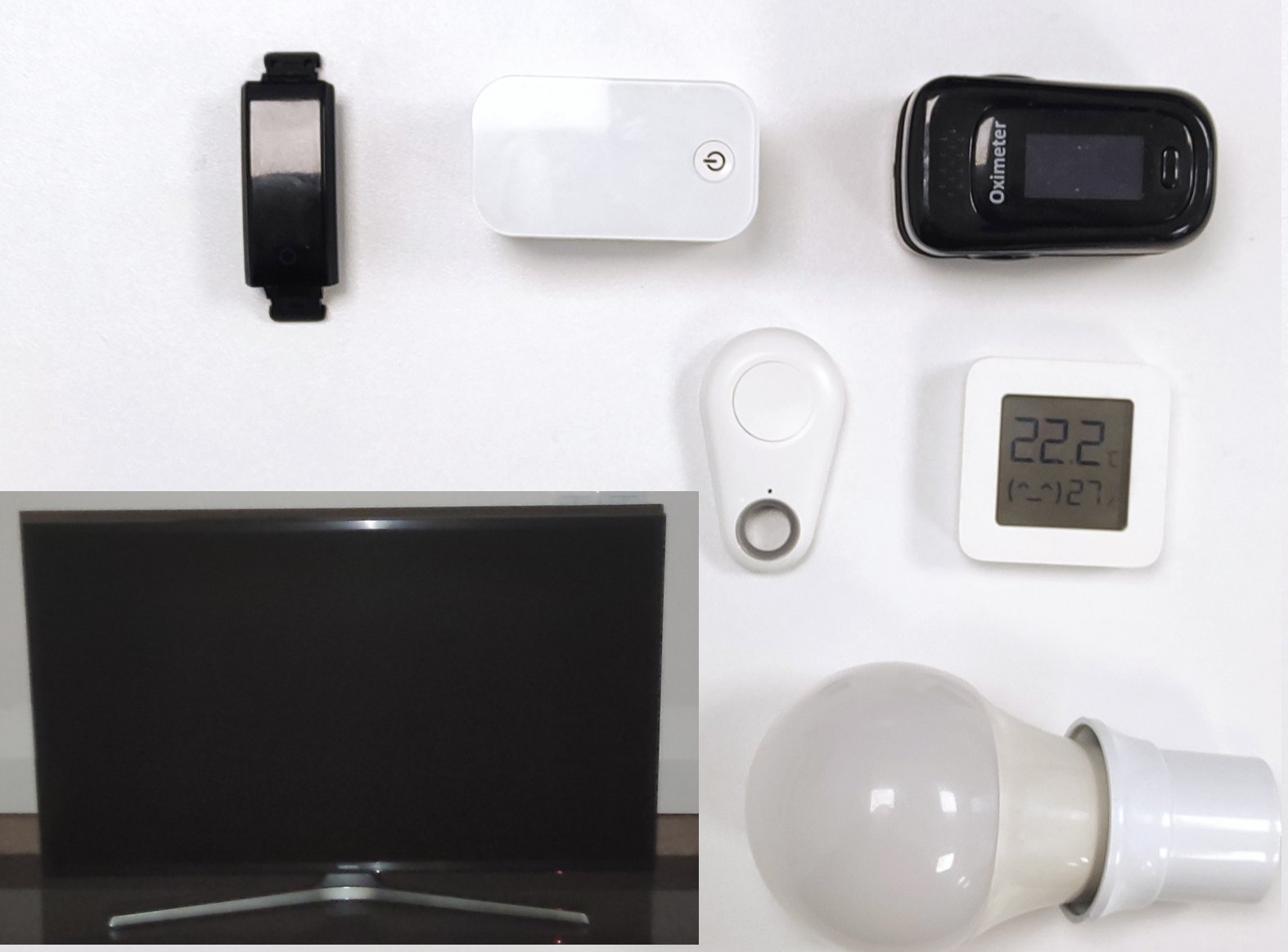}
\caption{BLE devices used in experiments}
\label{fig:testdevices}
\end{minipage}\hfill
\end{figure}

\subsubsection{Change Detection}
\begin{figure}[htp]
\centering
\includegraphics[width=6cm]{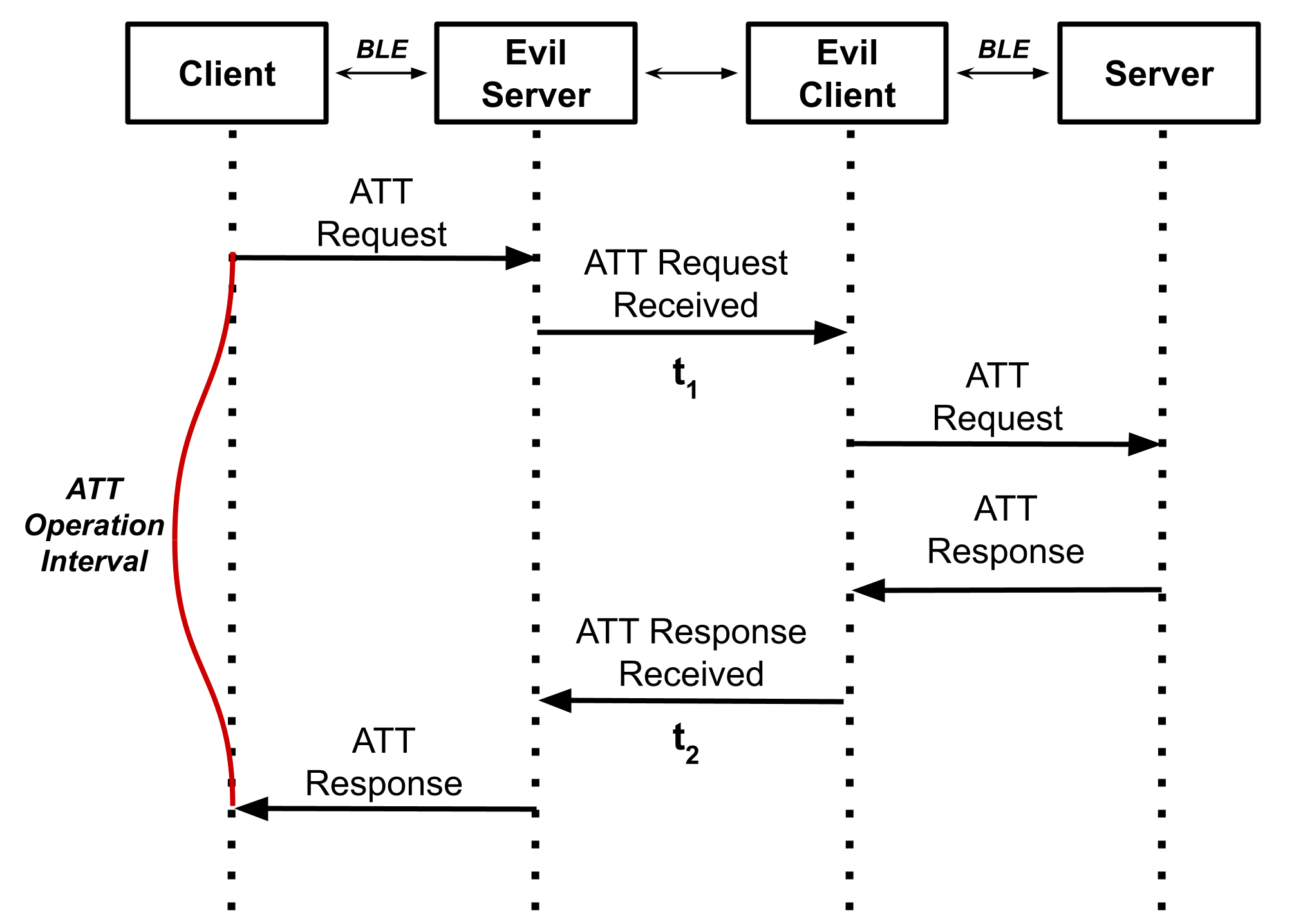}
\caption{ATT request-response traffic under an MITM attack.}
\label{fig:mitmatt}
\end{figure}

In a BLE MITM attack, the attacker acts as a proxy between the client and the server, and all involved ATT operations follow the steps shown in Fig. \ref{fig:mitmatt}. In this flow of events, the attacker adds two processing delay elements as denoted by $t_1$ and $t_2$ in Fig. \ref{fig:mitmatt}. 
Here, we ignore the over-the-air transmission time to and from the MITM attacker as it will be negligible in comparison to in-device processing time.

BLEKeeper attempts to establish a connection to each device identified as suspicious in order to detect potential spoofing attempts.
Once connected, it performs one of the select ATT operations on the device and the corresponding response time is measured.
Crucially, an MITM attack will only induce an increase in the response time.
Therefore, if no change is detected, the connection is terminated.
Otherwise, another ATT operation is performed to increase the likelihood of making a correct decision. 
The number of operations performed on the device can be set to obtain a target false-positive detection probability.

BLEKeeper measures the time interval between initial request and final response packets while performing ATT operation(s) as a client (shown on the left side of Fig. \ref{fig:mitmatt}).
In this regard, reliable detection of an attack depends on both the accuracy of the obtained device profile and having a model of the attacker-induced delay to the communication, defined as $T_{MITM}=t_1+t_2$.
The processing delay term $T_{MITM}$ due to an attack can be obtained through analysis of relevant functions in the codebase of publicly available tools. 
Since such an analysis will exclude other uncontrollable I/O related delays in the system, it will serve as a lower-bound estimate of $T_{MITM}$.

Given a BLE device's profile described in terms of the distribution of its response time values, 
$f^{i}_{\mathbf{Tr}}(Tr)$, and the measured minimum delay $T_{MITM}$ due to launching an MITM attack as a fixed value, our change detection method can be formulated as a simple hypothesis testing problem.
Accordingly, the goal of the detection system is to decide whether a measured response time is more likely to be generated due to a direct communication between BLEKeeper and server or due to an indirect communication through an attacker in between.
This can be realized by computing the likelihood ratio for a given response time measurement $Tr$ considering the two distributions that represents the device profile $P_i: f^{i}_{\mathbf {Tr}}(Tr)$ and its offsetted version with $T_{MITM}$, $P_{i,MITM}: f^{i}_{\mathbf{Tr}}(Tr)+T_{MITM}$.
Overall, an attack detection is made by deciding whether a measured ATT operation interval is more likely to be an instance of the former or the latter distribution.

\section{System Evaluation}

BLEKeeper has been developed with Python 3.7 and runs as a command line application on Raspberry Pi 4 Model B with 2 GB RAM. 
It uses three Ubertooth devices to sniff the BLE traffic in addition to the on-board Bluetooth interface (CYW43455) to perform ATT operations.
A picture of the system is shown in Fig. \ref{fig:detectionsystem}. 
To simulate the attacker, two Bluetooth CSR 4.0 dongles are used on a laptop with 16 GB RAM running Ubuntu 16.04 OS.
The system is evaluated using seven different BLE devices, shown in Fig. \ref{fig:testdevices}, and two attack tools with different architectures.

\begin{figure*}
	\centering
	\includegraphics[width=\textwidth]{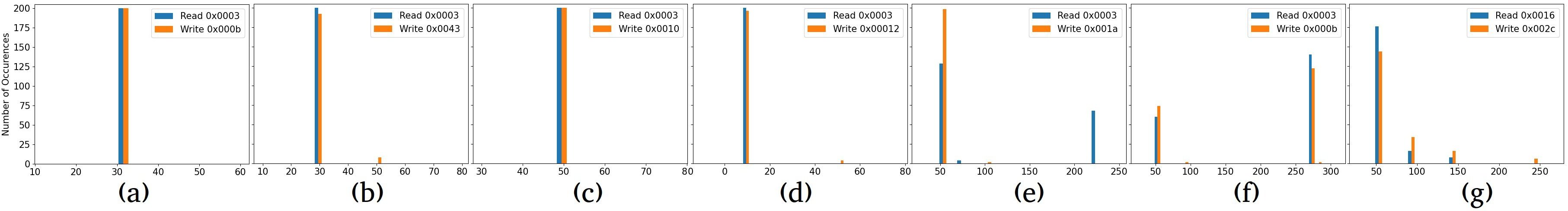}
	\caption{Distribution of response time values (ms) for (a) LEDBLE (b) LYWSD03MMC  (c) My Oximeter  (d) Medical  (e) MS1020   (f) iTag  (g) Samsung [TV].}
	\label{fig:testdeviceprofiles}
\end{figure*}

\subsection{Response Time Behavior of BLE Devices}

To profile BLE devices, 20 connections were established to each device to perform multiple ATT operations on both readable and writable characteristic values.
During each connection read and write operations were performed 10 times and response times were computed.
Table \ref{fig:deviceintervallist} provides mean and variance values corresponding to these measurements. 
As can be seen, response times vary from 10 and 300 milliseconds depending on the BLE device and the type of ATT operation.
More importantly, we observe that for four of the devices, response time values are very consistent, yielding a variance less than $10^{-5}$.
This indicates that for these devices a change in the response time can be used as a reliable attack indicator as the processing delay introduced by the MITM attacker is expected to be much higher.
Hence, for devices that exhibit a low variance, their response time behavior can be described by the mean value, $m_{i}$, instead of using the distribution, 
{\em i.e.,} $P_i: m_i$. 
However, for the remaining three devices, {\em i.e.,} MS1020, iTag, and Smart TV, the variation in measurements are in the order of several tens of milliseconds. 

Figure \ref{fig:testdeviceprofiles} presents the distribution of measured response time values for all devices.
It can be seen that for high variance devices, Fig. \ref{fig:testdeviceprofiles} (e)-(g), the distribution of values is not spread out but rather sharply clustered around a few discrete values, possibly indicating different modes of operation within the device.
Therefore, response time for those devices can be considered to follow a multimodal distribution consisting of two or three low-variance components.
In fact, 97\% of response times take one of two values for MS1020 and iTag and  one of three values for Samsung TV.
Correspondingly, for such devices, the device profile can be be expressed in terms of $j$ mean response time values, $\{m_{i,1},.., m_{i,j}\}$, corresponding to each cluster of measurements, {\em i.e.,} $P_i: \{m_{i,1},.., m_{i,j}\}$.

\begin{table*}[]
\caption{ Measured Response Time Values For Device Profiling and Attack Detection Performance}
\centering
\begin{tabular}{|c|c|c|c|c|c|c|c||c|c|}
\hline
\textit{\textbf{Device Name}}              & \textit{\textbf{Device Type}}       & \textit{\textbf{Operation}} & 
\textit{\textbf{Handle}} &
\textit{\textbf{Mean (sec)}} & \textit{\textbf{Variance (sec)}} & \textit{\textbf{Maximum (sec)}} & \textit{\textbf{Min (sec)}} & \textit{\textbf{FP (\%)}} & \textit{\textbf{FN (\%)}} \\ \hline
\multirow{2}{*}{LEDBLE}                    & \multirow{2}{*}{Smart Bulb}         & \textbf{Read}                  & 0x0003                & 0.03023                      & $ 9.72441*10^{-14} $                            & 0.030230                        & 0.030229                    & 0.00                 & 0.00                 \\ \cline{3-10} 
                                           &                                     & \textbf{Write}                 & 0x000b                & 0.03273                     & $ 1.28261*10^{-13} $                            & 0.032731                        & 0.032729                    & 0.00                 & 0.00                 \\ \hline
\multirow{2}{*}{LYWSD03MMC}                 & \multirow{2}{*}{Thermometer} & \textbf{Read}                  & 0x0003                & 0.03023                      & $ 6.08422*10^{-14} $                            & 0.030230                        & 0.030230                    & 0.00                 & 0.00                 \\ \cline{3-10} 
                                           &                                     & \textbf{Write}                 & 0x0043                & 0.03103                      & $ 1.53597*10^{-5} $                            & 0.050230                        & 0.030229                    & 4.00                 & 0.00                 \\ \hline
\multirow{2}{*}{My Oximeter}               & \multirow{2}{*}{Oximeter}           & \textbf{Read}                  & 0x0003                & 0.05008                      & $ 1.58237*10^{-7} $                            & 0.050230                        & 0.048979                    & 0.00                 & 0.00                 \\ \cline{3-10} 
                                           &                                     & \textbf{Write}                 & 0x0010                & 0.05023                      & $ 7.97839*10^{-11} $                            & 0.050294                        & 0.050229                    & 0.00                 & 0.00                 \\ \hline
\multirow{2}{*}{Medical}                   & \multirow{2}{*}{Oximeter}           & \textbf{Read}                  & 0x0003                & 0.01023                      & $ 7.46476*10^{-14} $                            & 0.010231                        & 0.010229                    & 0.00                 & 0.00                 \\ \cline{3-10} 
                                           &                                     & \textbf{Write}                 & 0x0012                & 0.01103                      & $ 3.14348*10^{-5} $                            & 0.050326                        & 0.010230                    & 2.00                 & 0.00                 \\ \hline
\multirow{2}{*}{MS1020}                    & \multirow{2}{*}{Smart Bracelet}     & \textbf{Read}                  & 0x0003                & 0.10998                      & $ 6.84789*10^{-3} $                            & 0.225229                        & 0.048979                    & 0.00                 & 0.00                 \\ \cline{3-10} 
                                           &                                     & \textbf{Write}                 & 0x001a                & 0.05073                      & $ 2.47513*10^{-5} $                            & 0.100231                        & 0.050229                    & 1.00                 & 0.00                 \\ \hline
\multirow{2}{*}{iTag}                      & \multirow{2}{*}{Smart Tag}          & \textbf{Read}                  & 0x0003                & 0.20647                      & $ 1.06311*10^{-2} $                            & 0.273978                        & 0.048978                    & 0.00                 & 0.00                 \\ \cline{3-10} 
                                           &                                     & \textbf{Write}                 & 0x000b                & 0.18908                      & $ 1.18340*10^{-2} $                            & 0.285228                        & 0.048978                    & 2.00                 & 0.00                 \\ \hline
\multirow{2}{*}{ Samsung {[}TV{]}} & \multirow{2}{*}{Smart TV}           & \textbf{Read}                  & 0x0016                & 0.05739                      & $ 5.21485*10^{-4} $                            & 0.146480                        & 0.048464                    & 4.00                 & 16.00                 \\ \cline{3-10} 
                                           &                                     & \textbf{Write}                 & 0x002c                & 0.07080                      & $ 1.81988*10^{-3} $                            & 0.243979                        & 0.048518                    & 3.00                 & 16.00                 \\ \hline
\multicolumn{6}{l}{\textbf{}}                                                                                                                                                                                                                                                                               & \multicolumn{1}{l|}{}           & \textbf{Average}            & \textbf{1.14}        & \textbf{2.28}        \\ \cline{8-10} 

\end{tabular}
\label{fig:deviceintervallist}
\end{table*}

\subsection{Characterizing MITM Attack Tools}

Launching an MITM attack requires an attacker to simultaneously operate as a BLE client and server. 
The attacker intercepts the data received in one BLE connection interface, modifies it when necessary, and relays it to the other interface for transmission.
There are several BLE MITM attack frameworks to perform these tasks. 
Most notably, these include \textit{BtleJuice} \cite{btlejuice}, \textit{Gattacker} \cite{gattacker}, and \textit{Mirage} \cite{mirage}.
\emph{BtleJuice} and \emph{Gattacker} require each Bluetooth connection to be controlled by an independent machine which communicates to each other using the TCP/IP stack. 
Unlike the other tools, \emph{Mirage} can simultaneously manage two interfaces on a single machine.
To characterize the amount of additional delay an MITM attack adds to response time, we examined the delay behavior of \emph{Btlejuice} and \emph{Mirage} attack tools.
(In our tests, we noticed that \emph{Gattacker} failed very frequently and decided not to include it in the subsequent measurements.)
To determine the minimum possible delay $T_{MITM}$ due to an MITM attack, we performed execution timing of these tools.
For this purpose, we identified functions involved in receiving, forwarding, and writing a packet.
Although these operations are computationally not very demanding, they nevertheless require switching context between user and kernel spaces several times.

\begin{table}[]
\caption{attack tool delay measurements}
\centering
\begin{tabular}{cc|c|c|c|c|}
\cline{3-6}
\multicolumn{1}{l}{}                                & \multicolumn{1}{l|}{}  & \multicolumn{2}{c|}{\textbf{$t_1$(ms)}}          & \multicolumn{2}{c|}{\textbf{$t_2$(ms)}}          \\ \hline
\multicolumn{1}{|c|}{\textit{\textbf{Device Name}}} & \textit{\textbf{Tool}} & \textit{\textbf{max}} & \textit{\textbf{min}} & \textit{\textbf{max}} & \textit{\textbf{min}} \\ \hline
\multicolumn{1}{|c|}{\multirow{2}{*}{LEDBLE}}       & Mirage                 & 9.8                   & 1.3                   & 9.0                   & \textbf{1.0}            \\ \cline{2-6} 
\multicolumn{1}{|c|}{}                              & Btlejuice              & 37.0                  & 6.0                   & 17.0                  & 2.0                   \\ \hline
\multicolumn{1}{|c|}{\multirow{2}{*}{LYWSD03MMC}}   & Mirage                 & 9.6                   & \textbf{1.2}          & 9.1                   & 1.0                     \\ \cline{2-6} 
\multicolumn{1}{|c|}{}                              & Btlejuice              & 36.0                  & 8.0                   & 15.0                  & 3.0                   \\ \hline
\end{tabular}
\label{attacktoolmeasurement}
\end{table}

To prevent server device related variations interfere with the measurements, we performed measurements using only the low-variance devices.
Table \ref{attacktoolmeasurement} provides measurements obtained while using the two attack tools to mount an MITM attack between two devices
during which 100 read and 100 write operations were performed.
These measurements essentially correspond to the $t_1$ and $t_2$ terms shown in Fig. \ref{fig:mitmatt}.
Overall, it is determined that the process of request and response forwarding takes on average 4.42 ms for \emph{Mirage} and 10.8 ms for \emph{Btlejuice}.
This difference can be attributed to \emph{Btlejuice} using two machines that communicate over TCP/IP.
More relevantly, the minimum value for the overall delay $T_{MITM}$ is measured to be 2.2 ms.

\subsection{Attack Detection Performance}

BLEKeeper listens to all advertising packets in an environment to identify servers through their device names and MAC addresses.
When BLEKeeper attempts to establish a connection with a server with identifier $i$, it retrieves its profile $P_i$ and measures a sequence of response times $Tr_1,\ldots, Tr_k$, where $k\geq1$, corresponding to each exchange between BLEKeeper and the server.  
Given a profile $P_i: \{m_{i,1}, \ldots, m_{i,j}\}$, comprising mean values associated with each of the $j$ components, 
the change detection algorithm determines a constellation of decision regions, $m_{i,j}-3\sigma_{i,j} \leq Tr \leq m_{i,j}+T_{MITM}$ where $T_{MITM}$ is set to 2.2
and $\sigma_{i,j}$ is the standard deviation computed around $m_{i,j}$.
Essentially, this decision region determines the range of response time values that reflects the normal behavior of the device. 
Then for all measurements $Tr_1,\ldots, Tr_k$ a decision is made as to whether a change due to an MITM attack is detected.  

In order to evaluate the success of BLEKeeper, MITM attacks are performed using both {\em Btlejuice} and {\em Mirage} tools. Response time measurements were obtained as BLEKeeper performed 50 read and 50 write operations on selected characteristics of server devices.
This yielded a total of 200 response time values for each device while under an attack.
The same measurements are repeated to obtain 200 additional response time measurements, corresponding to 100 read and write operations, without an attack.
Then using known device profiles and assuming $T_{MITM}=2.2$, a decision is made for each of the 400 response time values to test BLEKeeper's attack detection performance.
The last two columns of Table \ref{fig:deviceintervallist} provide resulting false-positive (FP) and false-negative (FN) detection rates for each device.

These results show that when low-variance devices are considered BLEKeeper is very effective in identifying all MITM attacks, producing a 0\% FN rate.
Surprisingly, though, we observed a 16\% FN rate with Smart TV. 
Given that the Smart TV's response time behavior exhibited the same clustering behavior as other devices, this result was not expected.
Examining the measurements, we determined that both attack tools yielded several delays around multiples of 50 ms.
We believe this is due to some queuing or context switching phenomenon in the host running attack tools. 
Samsung TV's response times, unlike other devices, are measured to cluster around  50, 100 and 150 ms. 
Due to this overlap, some of the values measured under attack essentially could not be distinguished from attack free measurements.
We also observed a FP rate of 1-4\% for devices that exhibit higher variance in ATT operations.
Better modeling of the distribution of response times, especially for the last three devices where the representation of device profiles, ({\em i.e.}, mean values of components) covered only 97\% of observed measurements, can further lower this rate.

Finally, we note that the above results are obtained when each response time $Tr_1,\ldots, Tr_k$ is evaluated individually. 
However, in practice, the measurements obtained from a given BLE connection must agree in the decision they yield. 
This can be utilized to further reduce both types of errors by requiring that a connection will be deemed under attack (or attack free) only if a number of measurements indicate so (or otherwise).
Considering a device with high FN rate, such as Samsung TV, the FN rate can be reduced dramatically by requiring at least half of the measurements to exhibit an anomalous change.

\section{Discussion and Conclusions}

We presented an active MITM attack detection system that identifies anomalous increases in the response time behavior of a BLE server when performing select ATT operations.
Our measurements most notably reveal that response time behavior of several BLE devices shows a strong regularity. 
We further observed that as low as 20 measurements are sufficient to obtain a reliable device profile, which can be performed in less than 20 seconds. We believe this behavior generalizes over a wide variety of BLE devices because of their simplicity.
We also demonstrate that response time behavior is a very reliable indicator for an MITM attack.
Our findings show that for devices that exhibit a low variance in their operation, a change in response time can 
detect the occurrence of an MITM attack without any errors even under a very conservative attack setting where the delay introduced by the attacker is assumed to be minimal.
Moreover, even for high variance devices that switch between discrete response time values, BLEKeeper yields a high detection accuracy. Overall, BLEKeeper can detect an attacker with 98\% accuracy by performing only a single read or write operation.

One limitation of our approach in the context of high variance devices is that an attacker may deliberately delay the response it received from the server for an appropriate duration before relaying to the client, thereby causing BLEKeeper to miss an attack.
One potential remedy to this behavior is for BLEKeeper to make a decision after performing a series of read and write operations.
Consistently delaying the response of the server will result in a more noticeable deviation in response time behavior as the device will appear to slow down.
Finally, we note that in order to identify a spoofing attempt, BLEKeeper's current setup requires a sniffer to track the connection and one BLE interface to perform the ATT operations. 
Ability to probe multiple servers simultaneously requires increasing the number of sniffers and BLE interfaces.

\bibliographystyle{plain}
\bibliography{bibliography}

\begin{thebibliography}{10}

\bibitem{assigned}
{Bluetooth SIG}.
\newblock Assigned numbers.
\newblock https://www.bluetooth.com/specifications/assigned-numbers.
\newblock Accessed Date: 2021-01-01.

\bibitem{bluetoothsig}
{Bluetooth SIG}.
\newblock {{Bluetooth} Market Update}.
\newblock https://www.bluetooth.com/bluetooth-resources/2020-bmu/.
\newblock Accessed Date: 2020-07-24.

\bibitem{bluetoothspec}
{Bluetooth SIG}.
\newblock {Core Specifications 5.2}.
\newblock
  https://www.bluetooth.com/specifications/bluetooth-core-specification/.
\newblock Accessed Date: 2020-08-05.

\bibitem{celosia2019fingerprinting}
Guillaume Celosia and Mathieu Cunche.
\newblock Fingerprinting {Bluetooth}-low-energy devices based on the generic
  attribute profile.
\newblock In {\em Proceedings of the 2nd International ACM Workshop on Security
  and Privacy for the Internet-of-Things}, pages 24--31, 2019.

\bibitem{btlejuice}
DigitalSecurity.
\newblock btlejuice.
\newblock https://github.com/DigitalSecurity/btlejuice.
\newblock Accessed Date: 2020-08-05.

\bibitem{galtier2020psd}
Florent Galtier, Romain Cayre, Guillaume Auriol, Mohamed Ka{\^a}niche, and
  Vincent Nicomette.
\newblock A psd-based fingerprinting approach to detect iot device spoofing.
\newblock In {\em 25th IEEE Pacific Rim International Symposium on Dependable
  Computing (PRDC 2020)}.

\bibitem{garbelini2020sweyntooth}
Matheus~E Garbelini, Chundong Wang, Sudipta Chattopadhyay, Sun Sumei, and
  Ernest Kurniawan.
\newblock Sweyntooth: Unleashing mayhem over {Bluetooth} low energy.
\newblock In {\em 2020 USENIX Annual Technical Conference (USENIX ATC 20)},
  pages 911--925, 2020.

\bibitem{gu2018bf}
Tianbo Gu and Prasant Mohapatra.
\newblock Bf-iot: Securing the iot networks via fingerprinting-based device
  authentication.
\newblock In {\em 2018 IEEE 15Th international conference on mobile ad hoc and
  sensor systems (MASS)}, pages 254--262. IEEE, 2018.

\bibitem{kang2018trusted}
James~Jin Kang, Kiran Fahd, and Sitalakshmi Venkatraman.
\newblock Trusted time-based verification model for automatic man-in-the-middle
  attack detection in cybersecurity.
\newblock {\em Cryptography}, 2(4):38, 2018.

\bibitem{lahmadi2020mitm}
Abdelkader Lahmadi, Alexis Duque, Nathan Heraief, and Julien Francq.
\newblock {MitM} attack detection in {BLE} networks using reconstruction and
  classification machine learning techniques.
\newblock In {\em MLCS 2020-2nd Workshop on Machine Learning for
  Cybersecurity}, 2020.

\bibitem{martin2019handoff}
Jeremy Martin, Douglas Alpuche, Kristina Bodeman, Lamont Brown, Ellis Fenske,
  Lucas Foppe, Travis Mayberry, Erik Rye, Brandon Sipes, and Sam Teplov.
\newblock Handoff all your privacy--a review of apple’s {Bluetooth} low
  energy continuity protocol.
\newblock {\em Proceedings on Privacy Enhancing Technologies}, 2019(4):34--53,
  2019.

\bibitem{crackle}
{mikeryan}.
\newblock Crackle.
\newblock https://github.com/mikeryan/crackle.
\newblock Accessed Date: 2020-08-05.

\bibitem{blurtooth}
{National Vulnerability Database}.
\newblock {CVE-2020-15802}.
\newblock https://cve.mitre.org/cgi-bin/cvename.cgi?name=CVE-2020-15802.
\newblock Accessed Date: 2020-09-14.

\bibitem{newaz2020survey}
AKM Newaz, Amit~Kumar Sikder, Mohammad~Ashiqur Rahman, and A~Selcuk Uluagac.
\newblock A survey on security and privacy issues in modern healthcare systems:
  Attacks and defenses.
\newblock {\em arXiv preprint arXiv:2005.07359}, 2020.

\bibitem{mirage}
{RCayre}.
\newblock mirage.
\newblock https://github.com/RCayre/mirage.
\newblock Accessed Date: 2020-08-05.

\bibitem{ryan2013bluetooth}
Mike Ryan.
\newblock {Bluetooth}: With low energy comes low security.
\newblock In {\em 7th USENIX Workshop on Offensive Technologies (WOOT 13)},
  2013.

\bibitem{gattacker}
{Securing}.
\newblock Gattacker.
\newblock https://github.com/securing/gattacker.
\newblock Accessed Date: 2020-08-05.

\bibitem{vallivaara2014detecting}
Visa~Antero Vallivaara, Mirko Sailio, and Kimmo Halunen.
\newblock Detecting man-in-the-middle attacks on non-mobile systems.
\newblock In {\em Proceedings of the 4th ACM conference on Data and application
  security and privacy}, pages 131--134, 2014.

\bibitem{wang2020bluedoor}
Jiliang Wang, Feng Hu, Ye~Zhou, Yunhao Liu, Hanyi Zhang, and Zhe Liu.
\newblock Bluedoor: breaking the secure information flow via {BLE}
  vulnerability.
\newblock In {\em Proceedings of the 18th International Conference on Mobile
  Systems, Applications, and Services}, pages 286--298, 2020.

\bibitem{wen2020firmxray}
Haohuang Wen, Zhiqiang Lin, and Yinqian Zhang.
\newblock Firmxray: Detecting {Bluetooth} link layer vulnerabilities from
  bare-metal firmware.
\newblock In {\em Proceedings of the 2020 ACM SIGSAC Conference on Computer and
  Communications Security}, pages 167--180, 2020.

\bibitem{wublueshield}
Jianliang Wu, Yuhong Nan, Vireshwar Kumar, Mathias Payer, and Dongyan Xu.
\newblock Blueshield: Detecting spoofing attacks in {Bluetooth} low energy
  networks.

\bibitem{yang2020beyond}
Jian Yang, Christian Poellabauer, Pramita Mitra, and Cynthia Neubecker.
\newblock Beyond beaconing: Emerging applications and challenges of {BLE}.
\newblock {\em Ad Hoc Networks}, 97:102015, 2020.

\bibitem{yaseen2019marc}
Muhammad Yaseen, Waseem Iqbal, Imran Rashid, Haider Abbas, Mujahid Mohsin,
  Kashif Saleem, and Yawar~Abbas Bangash.
\newblock Marc: A novel framework for detecting {MitM} attacks in ehealthcare
  {BLE} systems.
\newblock {\em Journal of Medical Systems}, 43(11):324, 2019.

\bibitem{255346}
Yue Zhang, Jian Weng, Rajib Dey, Yier Jin, Zhiqiang Lin, and Xinwen Fu.
\newblock Breaking secure pairing of {Bluetooth} low energy using downgrade
  attacks.
\newblock In {\em 29th USENIX Security Symposium (USENIX Security 20)}, pages
  37--54. {USENIX} Association, August 2020.

\end{thebibliography}

\vspace{12pt}

\end{document}